# NON-PERTURBATIVE EVALUATION

# OF THE

# PHYSICAL CLASSICAL VELOCITY

# IN THE

# LATTICE HEAVY QUARK EFFECTIVE THEORY

*Jeffrey E. Mandula*
*U.S. Department of Energy*
*Division of High Energy Physics*
*Washington, DC  20585, USA*

*and*

*Michael C. Ogilvie*
*Department of Physics*
*Washington University*
*St. Louis, MO  63130, USA*

# ABSTRACT

In the lattice formulation of the Heavy Quark Effective Theory, the value of the "classical velocity" $v$, as defined through the separation of the 4-momentum of a heavy quark into a part proportional to the heavy quark mass and a residual part which remains finite in the heavy quark limit ($P = Mv + p$), is different from its value as it appears in the bare heavy quark propagator ($S^{-1}(p) = v \cdot p$). The origin of the difference, which is effectively a lattice-induced renormalization, is the reduction of Lorentz (or O(4)) invariance to (hyper)cubic invariance. The renormalization is finite and depends specifically on the form of the discretization of the reduced heavy quark Dirac equation. For the Forward Time - Centered Space discretization, we compute this renormalization non-perturbatively, using an ensemble of lattices at $\beta = 6.1$ provided by the Fermilab ACP-MAPS Collaboration. The calculation makes crucial use of a variationally optimized smeared operator for creating composite heavy-light mesons. It has the property that its propagator achieves an asymptotic plateau in just a few Euclidean time steps. For comparison, we also compute the shift perturbatively, to one loop in lattice perturbation theory. The non-perturbative calculation of the leading multiplicative shift in the classical velocity is considerably different from the one-loop estimate, and indicates that for the above parameters, $\vec{v}$ is reduced by about 10-13%.

# I    INTRODUCTION

In the heavy quark limit[1,2] new dynamical symmetries emerge which give rise to a host of relations between decay constants and form factors of particles containing a heavy quark. For example, heavy quark spin-flavor symmetry implies that in the $M_Q \to \infty$ limit, a single form factor, the Isgur-Wise universal function $\xi$, describes all semileptonic decays of one meson containing a heavy quark into another, such as the process $B \to D^* l \nu$. From the first, it has been emphasized that although the heavy quark spin-flavor symmetry suffices to infer the existence of this function, its calculation requires non-perturbative techniques, such as lattice gauge theory[3]. Several such calculations have been carried out. A calculation by the present authors used a lattice implementation of the heavy quark effective theory[4], as did Hashimoto and Matsufuru[5] and the MILC collaboration[6], and several other lattice calculations have treated the heavy quarks as Wilson fermions with a small hopping constant, but avoided formally implementing the heavy quark limit[7,8].

On the lattice[4], as in the continuum[1,2,9], the Isgur-Wise limit entails the introduction of a "classical velocity" $v$, normalized to 1, which appears in the decomposition of the momentum of a heavy particle and in the reduced Dirac equation of the heavy quark field:

$$P = Mv + p$$
$$-iv \cdot D\, h^{(v)}(x) = 0 \qquad (1)$$

In the continuum, the velocity that appears in these two contexts is the same. However, as was first noted by Aglietti, on the lattice this is not the case[10]. This new renormalization is only possible because of the reduced symmetry of the lattice relative to continuum space-time.

To see that the classical velocity must be unchanged in the continuum, it is sufficient to realize that $v$ is the only 4-vector parameter in the heavy quark theory. Therefore, the vectors $v$ in the above two expressions must be proportional to each other, and since they are by definition both normalized to 1, they are equal. On the lattice, on the other hand, there are many linearly independent "4-vectors" that can be made from the components of $v$, where by a lattice 4-vector is meant a quantity that has the same transformation properties as $v$ under the lattice rotation-reflection group. The simplest examples are the vectors $(v_0^{2n+1}, v_1^{2n+1}, v_2^{2n+1}, v_3^{2n+1})$, each of whose components is the $(2n+1)^{st}$ power of the corresponding component of $v$. The renormalized lattice classical velocity can be proportional to any linear combination of these, subject only to an overall normalization condition. Since the possibility of a renormalization of the classical velocity arises purely because of the discretization of space-time, the actual shift is very sensitive to the details of the how the heavy quark theory is implemented on the lattice.

In this paper we describe a non-perturbative calculation of the classical velocity shift. It is based on a computation of the shift in the energy of a composite meson containing one heavy and one light quark, as measured by the change in the rate of fall-off of its Euclidean space propagator, for a given shift in its residual momentum. The simulation was performed on an ensemble of $24^3 \times 48$ quenched lattices at $\beta = 6.1$ provided by the Fermilab ACP-MAPS collaboration. Our analysis involved two novel techniques which we will describe in detail. One was to systematically expand all quantities in the simulation in powers of the space components of $v$. This exploits the structure of our discretization of the Dirac equation, which has the property that after n lattice time steps, the heavy quark propagator is an $(n-1)^{st}$ order polynomial in the components of $\tilde{v} \equiv \vec{v}/v_0$. Thus the structure of Aglietti's effective theory for slow quarks[11] is intrinsic on a finite lattice. The principal advantage of using this expansion is that it makes clear precisely what information can be extracted from the simulation, and does so with the minimum of computation. In this calculation we also used a variationally optimized smeared operator for creating composite heavy-light mesons, following the ideas of Draper, McNeile, and Nenkov[12]. The propagator of the optimized field achieves an asymptotic plateau in just a few Euclidean time steps. Its use allows us to achieve stable and reliable results from a relatively modest number of lattices.

For comparison, we also carried out a one-loop perturbative calculation of the renormalization of the classical velocity, following the analyses of Aglietti[10] and Aglietti and Giménez[13]. Those authors, however, used a different discretization of the lattice Dirac operator from that we have found convenient to use in simulations, and so the results of our calculations are different.

This paper is organized as follows. In Sec. II we briefly review the lattice heavy quark effective theory and introduce the expansion in powers of the classical velocity. In Sec. III we show how the physical value of the classical velocity can be found from the rate of fall-off of a composite particle propagator, and express its dependence on the input classical velocity in terms of quantities computed directly in the expansion in powers of $\tilde{v}$. In Sec. IV we describe the procedure for choosing a composite meson field which is variationally optimized to create from the space of states of a given classical velocity that state with as large an overlap with the ground state as possible. We show how quickly the ground state saturates the propagator depends on the space of trial states. In Sec. V we show the results of a Monte Carlo simulation implementing this analysis and obtain the leading multiplicative correction to the classical velocity. In Sec. VI we carry out the one-loop perturbative calculation of the classical velocity renormalization and compare the results to the simulation and to the calculations of Aglietti and Aglietti and Giménez. We conclude with some summary remarks in Sec VII.

## II   THE LATTICE HEAVY QUARK EFFECTIVE THEORY – EXPANSION IN $\tilde{v}$

The incorporation of the heavy quark limit into a Lagrangian defines the heavy quark effective theory (HQET) and is accomplished by factoring out a phase which is singular in the $M \to \infty$ limit and defining a reduced field[9]

$$\frac{1 + \gamma \cdot v}{2} h^{(v)}(x) = \lim_{M \to \infty} e^{-iMv \cdot x} \frac{1 + \gamma \cdot v}{2} \psi(x) \qquad (2)$$

In the limit, the Lagrangian for $h^{(v)}(x)$ becomes

$$\mathcal{L}^{(v)} = \bar{h}^{(v)}(x) \, iv \cdot D \, h^{(v)}(x) = \bar{h}^{(v)}(x) \, iv \cdot [\partial - igA(x)] \, h^{(v)}(x) \qquad (3)$$

The Dirac equation for the reduced field that follows from this Lagrangian is the starting point for this discussion, Eq. (1). There is a separate field for each value of the classical velocity and for each heavy particle and antiparticle type. Note that the reduced field has only one component; the spinor structure of its propagator is a fixed matrix, the projection operator on positive energy states with velocity $v$. Heavy quark loops arise only in higher orders in $1/M$, and in this paper we work exclusively in the limiting theory.

Formulating the HQET on the lattice first requires transcribing it to Euclidean space. In doing this, one must remember that the classical velocity is a fixed external parameter, and so its time component does not change when time and energy are analytically continued to imaginary values. Therefore, when after the Wick rotation we adopt a Euclidean coordinate system, in that coordinate system, the classical velocity's time component is imaginary.

$$v = (iv_0, \vec{v})$$

$$v_0 = \sqrt{1 + \vec{v}^2} \tag{4}$$

Also, since the conventional Minkowski and Euclidean space metrics are opposite ($x_0^2 - x_1^2 - x_2^2 - x_3^2$ vs $x_0^2 + x_1^2 + x_2^2 + x_3^2$), the Euclidean space normalization of $v$ is

$$v^2 = -1 \tag{5}$$

The lattice heavy quark effective theory is effectively defined by a choice of discretization of the reduced Dirac equation. Denoting the propagator of the reduced field $h^{(v)}(x)$ by $S^{(v)}(x,y)$, we choose a discretization which is symmetrical in the spatial directions, but uses an asymmetric, forward difference in the time direction:

$$v_0 [ U_4(x,x+\hat{t}) S^{(v)}(x+\hat{t},y) - S^{(v)}(x,y) ]$$
$$+ \sum_{\mu=1}^{3} \frac{-iv_\mu}{2} [U_\mu(x,x+\hat{\mu}) S^{(v)}(x+\hat{\mu},y) - U_\mu(x,x-\hat{\mu}) S^{(v)}(x-\hat{\mu},y)] = \delta(x,y) \tag{6}$$

We have described elsewhere some of the structural aspects of the this theory, especially the manner in which the effects of the "fermion doubler" states are suppressed[14].

The use of an asymmetric time difference is not optional. The reason is that in the $M \to \infty$ limit the quarks only propagate in one direction of time. With that boundary condition, if a symmetrically discretized first time difference were used, the resulting heavy quark propagator would not have a continuum limit, because it would vanish on alternate sites in the positive time direction[15]. The requirement that heavy quarks propagate only forward in time means that the heavy quark propagator is obtained by forward recursion. No relaxation methods are needed, and so simulations with the lattice heavy quark effective theory involve only modest computational needs.

If we divide through the lattice Dirac equation by $v_0$, we see that the $i^{th}$ component of the normalized velocity vector, $\tilde{v}_i = v_i/v_0$ plays the role of a transverse hopping constant. That is, the reduced heavy quark propagator gets one factor of $\tilde{v}_i$ for each unit displacement in the $\pm\ i^{th}$ direction. Since the propagator at any particular site on the $n^{th}$ time slice depends only on its value at the same spatial site or sites displaced by one transverse lattice spacing on the previous time slice, after n time steps the propagator is an $(n-1)^{st}$ order polynomial in the components $\tilde{v}_i$.

The calculation of the heavy quark propagator in simulations is greatly facilitated by exploiting this structure and expanding the propagator in a power series in $\tilde{v}_i$:

$$S^{(v)}(t,\vec{x}) = \sum_{m_1,m_2,m_3} \tilde{v}_1^{m_1} \tilde{v}_2^{m_2} \tilde{v}_3^{m_3} S(t,\vec{x},\vec{m}) \qquad (7)$$

The site $\vec{x}$ is measured from the starting location. The computation of the coefficients in this polynomial is highly efficient. The value of the index $m_i$ is the maximum transverse lattice displacement of the heavy quark propagator in the $i^{th}$ direction contributing to that coefficient. More precisely, the coefficient $S(t,\vec{x},\vec{m})$ is non-zero only on the sites

$$x_i = -m_i, -m_i+2, \ldots, +m_i \qquad (i = 1,2,3)$$

$$\sum_{i=1}^{3} m_i \leq t - 1 \tag{8}$$

Since each transverse hop in the computation of $S(t,\vec{x},\vec{m})$ can occur at any time between the initial and final times, the relative growth of these coefficients with $t$ will have a factor

$$S(t,\vec{x},\vec{m}) \propto t^{m_1 + m_2 + m_3} \tag{9}$$

It will prove very efficacious to systematically expand all quantities that depend on the heavy quark propagator in analogous power series in the normalized classical velocity.

## III    THE PHYSICAL CLASSICAL VELOCITY

The physical classical velocity of a particle is straightforwardly defined by the division of a heavy particle's 4-momentum into a fixed part proportional to the heavy quark mass and a residual term which is finite in the $M \to \infty$ limit, the first expression in Eq. (1). The particle can be either a heavy quark or a composite object containing a heavy quark, whose classical velocity is the same a that of its heavy quark component. With this definition[*] of $v^{(phys)}$, the physical residual energy of a particle containing a heavy quark, relative to the energy of its quark component, is given by

$$E^{(v)}(\vec{p}) = \lim_{M \to \infty} \sqrt{(M+m)^2 + ((M+m)\vec{v}^{(phys)} + \vec{p})^2} - M v_0^{(phys)}$$

$$= m v_0^{(phys)} + \tilde{v}^{(phys)} \cdot \vec{p} \qquad ( \tilde{v}^{(phys)} \equiv v_i^{(phys)}/v_0^{(phys)} )$$

(10)

In this expression, $m$ is the difference between the composite particle mass and the heavy quark mass $M$. This is of course a continuum relation. On an infinite lattice, the shift in the classical velocity can have additional terms which are higher order in the residual 3-momentum. Nonetheless, we may take as a definition of the normalized physical classical velocity $\tilde{v}^{(phys)}$ the coefficient of the leading (linear) term in an expansion of the residual energy of a heavy particle in its residual 3-momentum, or equivalently its derivative with respect to $\vec{p}$ at $\vec{p} = 0$:

---

[*]   Throughout this paper we reserve the symbol $v$ with no $^{(phys)}$ superscript for the input, or bare classical velocity.

$$\tilde{v}^{(phys)} = \left.\frac{\partial E^{(v)}(\vec{p})}{\partial \vec{p}}\right|_{\vec{p}=0} \tag{11}$$

In simulations the function $E^{(v)}(\vec{p})$ can be found from the rate of falloff of a heavy particle's propagator. In perturbation theory it can be found from the shift of the pole of the residual propagator in the complex residual energy plane. Equation (11) is the basis of our simulation of the physical classical velocity.

In simulations the lattices are of course not infinite, and this can be expected to introduce errors in the determination of the classical velocity shift. Like all finite-size errors these can in principal only be eliminated by extrapolation to the infinite volume limit. However, from the foregoing discussion of the structure of the lattice residual Dirac equation, it is clear that if $m_i$ is not too large, the only finite volume errors in the coefficients $S(t,\vec{x},\vec{m})$ are those induced by the finite-volume errors already present in the link variables. Specifically, for $m_i$ less than half the number of lattice sites on a side (reduced by the smearing size for composite particle propagation), those coefficients develop no further finite-volume errors because the limit on the number of transverse hops means that the propagator is not influenced by the lattice at distances more than $\pm m_i$ sites in the i$^{th}$ direction from the starting position.

We may apply the above definition of the physical classical velocity to either the propagator of a heavy quark or to the propagator of a heavy-light composite particle. In perturbation theory it is most natural to directly apply it to the residual heavy quark propagator, but in simulations the other choice is both more natural and more convenient. In order to simulate the heavy quark propagator, we would have to apply a global gauge-fixing procedure just to make it non-zero. Furthermore, in order to have the previous symmetry discussion remain valid, we would have to choose a (lattice) covariant gauge, and since the asymptotic falloff of the propagator is needed, the gauge would have to be "smooth". Presumably a gauge such as the lattice Landau gauge would be adequate. In perturbation theory this is all accomplished by the choice of the gluon propagator, and so it is essentially automatic. Non-perturbatively this step must be carried out explicitly, and it would demand rather more computation than the calculation of the heavy quark propagator itself.

In the following we will simulate the physical classical velocity of a meson made of one heavy and one light quark. Its propagator is a physical quantity, amenable to direct simulation without global gauge fixing. Its classical velocity will be the same as that of its heavy quark component. This is because the mass difference between the heavy quark and the composite meson is finite, which corresponds to a finitely different breakup of the total 4-momentum

$$P = Mv + p \rightarrow (M + m)v' + p' \tag{12}$$

In the heavy quark limit, both $v - v'$ and $p - p'$ vanish like $m/M$.

We expand the composite meson propagator in a power series in the input classical velocity, parallelling the expansion of the heavy quark propagator, Eq. (7).

$$M^{(v)}(t,\vec{p}) = \sum_{m_1,m_2,m_3} \tilde{v}_1^{m_1} \tilde{v}_2^{m_2} \tilde{v}_3^{m_3} M(t,\vec{p},\vec{m}) \tag{13}$$

Since the classical velocity enters only through the heavy quark propagator, each coefficient in this series for the meson propagator involves only one of the coefficients in the heavy quark propagator series, that with just the same value of $\vec{m}$.

The gauge invariant propagator $M^{(v)}$ satisfies lattice rotation-inversion symmetry conditions. For a scalar meson, the propagator is invariant under simultaneous lattice transformations of $\vec{v}$ and $\vec{p}$. To express this in terms of the coefficients $M(t,\vec{p},\vec{m})$, recall that the 48 element 3-dimensional lattice rotation-reflection group is equivalent to the group of permutations on the three axes times the three independent inversion groups of each axis[16]. The coefficients are then invariant under simultaneous permutations of the components of $\vec{p} = (p_1, p_2, p_3)$ and $\vec{m} = (m_1, m_2, m_3)$. In addition, $M(t,\vec{p},\vec{m})$ is an even or an odd function of $p_i$ depending on whether $m_i$ is an even or an odd integer, respectively.

For fixed $\tilde{v}$ and $\vec{p}$, the asymptotic behavior of the propagator is

$$M^{(v)}(t,\vec{p}) \sim C^{(v)}(\vec{p}) e^{-E^{(v)}(\vec{p})t} \qquad (14)$$

We extract the physical classical velocity $\tilde{v}^{(phys)}$ from $M^{(v)}$ by taking its logarithmic derivative with respect to $\vec{p}$, evaluated at zero momentum.

$$\begin{aligned}\frac{\partial M^{(v)}(t,\vec{p})/\partial p_i |_{\vec{p}=0}}{M^{(v)}(t,\vec{p}=0)} &\sim \frac{\partial C^{(v)}(\vec{p})/\partial p_i |_{\vec{p}=0}}{C^{(v)}(\vec{p}=0)} - \left.\frac{\partial E^{(v)}(\vec{p})}{\partial p_i}\right|_{\vec{p}=0} t \\ &= \frac{\partial C^{(v)}(\vec{p})/\partial p_i |_{\vec{p}=0}}{C^{(v)}(\vec{p}=0)} - \tilde{v}_i^{(phys)} t \end{aligned} \qquad (15)$$

The basic elements of the simulation are the coefficients $M(t,\vec{p},\vec{m})$ which appear in the expansion of $M^{(v)}$ in powers of the input classical velocity. Therefore we develop the logarithmic derivative in an analogous series in the input classical momentum $\tilde{v}$ :

$$\frac{\partial M^{(v)}(t,\vec{p})/\partial p_i \big|_{\vec{p}=0}}{M^{(v)}(t,\vec{p}=0)} \;=\; \sum \tilde{v}_1^{m_1}\,\tilde{v}_2^{m_2}\,\tilde{v}_3^{m_3}\, R^{(i)}(t,\vec{m}) \tag{16}$$

Each of the expansion functions $R^{(i)}(t,\vec{m})$ is asymptotically linear in $t$, and the negative of its slope is the corresponding term in the expansion for $\tilde{v}_i^{(phys)}$:

$$\begin{aligned} R^{(i)}(t,\vec{m}) &\sim\; const \;-\; c^{(i)}(\vec{m})\, t \\ \tilde{v}_i^{(phys)} &=\; \tilde{v}_1^{m_1}\,\tilde{v}_2^{m_2}\,\tilde{v}_3^{m_3}\, c^{(i)}(\vec{m}) \end{aligned} \tag{17}$$

To adapt this discussion to the actual situation of finite-extent lattices, the only modification required is the replacement of the continuum momentum derivative by a lattice approximation, which we take to be the symmetrical first difference on the finite Fourier transform lattice.

$$\begin{aligned} \frac{\partial M^{(v)}(t,\vec{p})}{\partial p_i}\bigg|_{\vec{p}=0} &\Rightarrow\; \frac{\Delta M^{(v)}(t,\vec{p})}{\Delta p_i}\bigg|_{\vec{p}=0} \\ &\equiv\; \frac{1}{2 p_{min}} \left[ M^{(v)}(t,\, p_i = p_{min}) - M^{(v)}(t,\, p_i = -p_{min}) \right] \end{aligned} \tag{18}$$

where $p_{min} = 2\pi/Na$ is the smallest finite momentum representable on a lattice with spacing $a$ and $N$ sites to a side. Note that because of the lattice symmetries, the derivative or symmetrical first difference of $M^{(v)}(t,\vec{p})$ with respect to $p_i$ at $\vec{p}=0$ must be odd in $\tilde{v}_i$ and even in the orthogonal components $\tilde{v}_j$ ($j \ne i$).

The shift in the classical velocity to a given order $\vec{m}$ in the components of the bare classical velocity only depends on the heavy quark propagator at values of the spatial coordinates up to $\vec{m}$ lattice sites away from the initial heavy quark location. Explicitly, for the heavy quark residual propagator,

$$\left.\frac{\Delta S^{(v)}(t,\vec{p})}{\Delta p_i}\right|_{\vec{p}=0, O(v_i^n)} = \frac{Na}{4\pi} \sum_{|x_i| \leq n} \left(2i \sin \frac{2\pi x_i}{N}\right) S^{(v)}(t,\vec{x}) \qquad (19)$$

By expanding the composite particle propagator and its momentum derivative or difference in the logarithmic derivative Eq. (15) in powers of $\tilde{v}$, Eq. (13), we can identify the coefficient functions $R^{(i)}(t,\vec{m})$. The non-vanishing terms through 3$^{rd}$ order are:

$$\begin{aligned}
R^{(i)}(t,m_i=1,m_{j\neq i}=0) &= \frac{\Delta M(t,\vec{p},m_i=1,m_{j\neq i}=0)/\Delta p_i|_{\vec{p}=0}}{M(t,\vec{p}=0,\vec{m}=0)} \\
R^{(i)}(t,m_i=3,m_{j\neq i}=0) &= \frac{\Delta M(t,\vec{p},m_i=3,m_{j\neq i}=0)/\Delta p_i|_{\vec{p}=0}}{M(t,\vec{p}=0,\vec{m}=0)} \\
&\quad - \frac{\Delta M(t,\vec{p},m_i=1,m_{j\neq i}=0)/\Delta p_i|_{\vec{p}=0}\; M(t,\vec{p}=0,m_i=2,m_{j\neq i}=0)}{M(t,\vec{p}=0,\vec{m}=0)^2} \\
R^{(i)}(t,m_i=1,m_{j\neq i}=(2,0)) &= \frac{\Delta M(t,\vec{p},m_i=1,m_{j\neq i}=(2,0))/\Delta p_i|_{\vec{p}=0}}{M(t,\vec{p}=0,\vec{m}=0)} \\
&\quad - \frac{\Delta M(t,\vec{p},m_i=1,m_{j\neq i}=0)/\Delta p_i|_{\vec{p}=0}\; M(t,\vec{p}=0,m_i=2,m_{j\neq i}=0)}{M(t,\vec{p}=0,\vec{m}=0)^2}
\end{aligned} \qquad (20)$$

Through third order all others vanish by lattice symmetry. In simulations, one can of course further exploit the lattice symmetries by averaging over directions to improve signal-to-noise ratios.

# IV  VARIATIONAL OPTIMIZATION OF THE HEAVY-LIGHT MESON OPERATOR

It is well known that a gauge-invariant heavy-light meson field consisting of component fields at the same point in space-time is a poor choice in simulations, because many time steps are needed for the excited meson states to die off and for the ground state to dominate its propagator. While any smeared operator containing the reduced heavy quark field can be expected to improve the rate of convergence, in actual simulations of quantities defined via composite fields, the limiting factor in the precision of the final result is usually how quickly the ground state in a given sector dominates the propagator. This is so because the statistical precision of propagators deteriorates rapidly with increasing time, and so it is crucial that the propagator reach its asymptotic form in as few time steps as possible.

We implement this requirement variationally, adapting the procedure of Draper, McNeile, and Nenkov[12]. In any sector defined by a given set of quantum numbers and a residual momentum, the ground state contribution to a composite field propagator has the slowest rate of decay. Therefore, the requirement that the composite field be chosen so that its propagator reaches its asymptotic form in as few time steps as possible is equivalent to the requirement that the state created by the composite field have as large a component as possible along the ground state. This in turn is equivalent to the requirement that for any given time separation, the composite field is chosen so that the magnitude of its propagator is as large as possible.

Specifically, we consider composite operators that, in the Coulomb gauge, as functions of time and residual momentum have the form

$$\Psi^{(v)}(t,\vec{p}) = \sum_{\vec{y}} \psi^{(v)}_{[\vec{p}]}(\vec{y}) \left[ \sum_{\vec{x}} e^{-i\vec{p}\cdot\vec{x}} q(x) h^{(v)}(x+y) \right] \qquad (21)$$

where $q(x)$ is a light quark field, $\psi$ is a relative coordinate weighting function and the sum goes over sites $y$ on the same time slice as $x$. For $\vec{v} = 0$, $\psi$ is effectively the wave function of the composite meson, but in general the weighting function does not have this interpretation. The propagator of this composite field is

$$M^{(v)}_{[\psi]}(t,\vec{p}) = \sum_{\vec{y},\vec{y}'} \psi^{(v)*}_{[\vec{p}]}(\vec{y}) K^{(v)}_{[\vec{p}]}(t,\vec{y},\vec{y}') \psi^{(v)}_{[\vec{p}]}(\vec{y}') \qquad (22)$$

where

$$K^{(v)}_{[\vec{p}]}(t,\vec{y},\vec{y}') = \sum_{\vec{x}} e^{-i\vec{p}\cdot\vec{x}} \left\langle s(x,x'=0) S^{(v)}(x+y,y') \right\rangle \qquad (23)$$

The average is over the ensemble of lattices. It is amusing to note that while it might seem intuitive to take the location of the heavy quark as nominal coordinate of the meson field, it is handiest to use the location of the light quark as the composite particle's coordinate. The requirement that the normalized propagator be maximal on a given time slice implies that $\psi^{(v)}$ is that eigenvector of $K^{(v)}$ with the largest eigenvalue:

$$\sum_{y'} K^{(v)}_{[p]}(t,y,y') \psi^{(v)}_{[p]}(t,y') = \lambda^{(v)}_{[p]}(t) \psi^{(v)}_{[p]}(t,y) \qquad (24)$$

The largest eigenvalue is, in fact, the value of the optimal meson propagator on time slice t.

$$M^{(v)}_{[\psi_{(opt)}]}(t,\vec{p}) = \lambda^{(v)}_{[\vec{p}]}(t) \qquad (25)$$

Note that in this specification of the maximization, there is an independent variational condition for each value of $t$. As we shall see, this is not a redundancy (except for $\vec{v} = 0$), but is needed to represent the actual structure of the weighting function.

The solution to this eigenvalue problem is facilitated by expanding the kernel, eigenvalue, and eigenvector in power series in the components of the classical velocity $\tilde{v}_i$, exactly as in the propagator expansions.

$$\left\{ K^{(v)}_{[\vec{p}]}(t,\vec{y},\vec{y}'), \lambda^{(v)}_{[\vec{p}]}(t), \psi^{(v)}_{[\vec{p}]}(t,\vec{y}) \right\}$$
$$= \sum_{m_1,m_2,m_3} \tilde{v}_1^{m_1} \tilde{v}_2^{m_2} \tilde{v}_3^{m_3} \left\{ K_{[\vec{p}]}(t,\vec{y},\vec{y}',\vec{m}), \lambda_{[\vec{p}]}(t,\vec{m}), \psi_{[\vec{p}]}(t,\vec{y},\vec{m}) \right\} \qquad (26)$$

The problem then takes the structure of the perturbative analysis of a general eigenvector/eigenvalue problem. In particular, it is only the static, zeroth order equation that requires finding the eigenvalues of a matrix, and it is only the zeroth order kernel matrix that must be inverted. Both the final time $t$ and the residual momentum of the meson $\vec{p}$ are fixed parameters.

The dependence of the zeroth order kernel on the residual momentum is very simple. Once the $\vec{p} = 0$ problem is solved, the results for non-zero $\vec{p}$ can be immediately read off. The zeroth order kernel

$$K_{[\vec{p}]}(t,\vec{y},\vec{y}',\vec{m}=0) = \sum_{\vec{x}} e^{-i\vec{p}\cdot\vec{x}} \left\langle s(x,x'=0) S(x+y,y',\vec{m}=0) \right\rangle \qquad (27)$$

simplifies because, to that order the residual heavy quark propagator has no transverse hops. Thus the sum over $\vec{x}$ has only one contribution, at $\vec{x} = \vec{y}' - \vec{y}$, which displays the momentum dependence as a diagonal similarity transformation:

$$K^{(0)}_{[\vec{p}]}(t,\vec{y},\vec{y}') = e^{i\vec{p}\cdot\vec{y}} \left\langle s(\vec{y}'- \vec{y}, \vec{x}'= 0) S^{(v)}(t,\vec{y}';0\vec{y}') \right\rangle e^{-i\vec{p}\cdot\vec{y}'} \tag{28}$$

The zeroth order eigenvalue is thus independent of the residual momentum, as it should be, and the residual momentum dependence of the zeroth order eigenvector is simply a phase.

$$\psi_{[\vec{p}]}(t,\vec{y},\vec{m}=0) = e^{i\vec{p}\cdot\vec{y}}\psi_{[0]}(t,\vec{y},\vec{m}=0)$$
$$\lambda_{[\vec{p}]}(t,\vec{m}=0) = \lambda_{[0]}(t,\vec{m}=0) \tag{29}$$

The zeroth order term in the eigenvalue equation Eq. (24) at $\vec{p} = 0$ is the starting point for its solution:

$$\sum_{\vec{y}'} K_{[0]}(t,\vec{y},\vec{y}',\vec{m}=0) \psi_{[0]}(t,\vec{y}',\vec{m}=0) = \lambda_{[0]}(t,\vec{m}=0) \psi_{[0]}(t,\vec{y},\vec{m}=0) \tag{30}$$

We find the eigenvalues and eigenvectors of this equation numerically, over the space of functions which are invariant under the lattice cubic rotation/reflection group and non-vanishing on the sites contained in a box with $(2N_w + 1)$ sites on a side. Symmetries reduce the effective dimension of the matrix equation from $(2N_W + 1)^3$ to $(N_W + 1)(N_W + 2)(N_W + 3)/6$. As noted above, the eigenfunction for non-zero $\vec{p}$ is obtained from this by multiplying by the appropriate phase, Eq. (29).

The quality of the solution improves rapidly with increasing $N_W$. Figure 1 shows the effective apparent mass, given from the rate of fall-off from one time slice to the next, for a range of domains over which the eigenvalue problem is solved ranging from $N_W = 0$ to $N_W = 6$. The error bars are full jackknife statistical errors from the ensemble of 42 lattices, and the curves are from two-exponential fits to the underlying propagators.

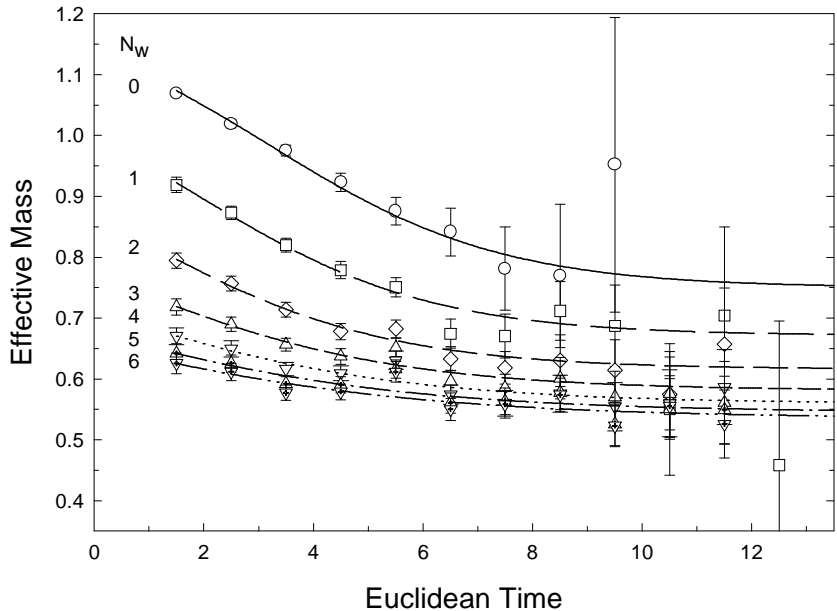

Figure 1 — Convergence to an effective mass plateau of a static heavy-light meson propagator optimized within domains containing $(2N_W + 1)^3$ sites.

Several remarks are in order regarding these results. The first thing to note is that the local ($N_W = 0$) operator never reaches a plateau. That is, without some sort of smearing procedure for enhancing the convergence of the meson propagator, this program for non-perturbatively computing the renormalization of the classical velocity would have failed. Another observation is that it is the parallel undulations of the data points for the different sized domains reflect correlations from one time slice to the next within the ensemble of lattices. It is the magnitude of these undulations, not the statistics of each data point, that is a meaningful measure of the precision with which the effective mass is determined. To within that precision, no evident plateau is found for $N_W = 0$ or for $N_W = 1$ (27 sites). For $N_W = 2$ (125 sites), a plateau is reached by about t = 4. For the larger domains, $N_W > 2$ ($7^3$ to $13^3$ sites), a plateau is evident by time slice t = 3. In fact, given the limited precision indicated by the parallel undulations, one cannot be sure that a plateau is not reached even more quickly. The fact that the successive fits asymptote to progressively lower masses is an artifact of the fitting procedure. Even with ideal data, a two-exponential fit will overestimate the asymptotic mass. The overestimate becomes smaller as more of the higher mass (subdominant) contributions to the propagator are removed, but remains finite so long as any high-mass contributions are present.

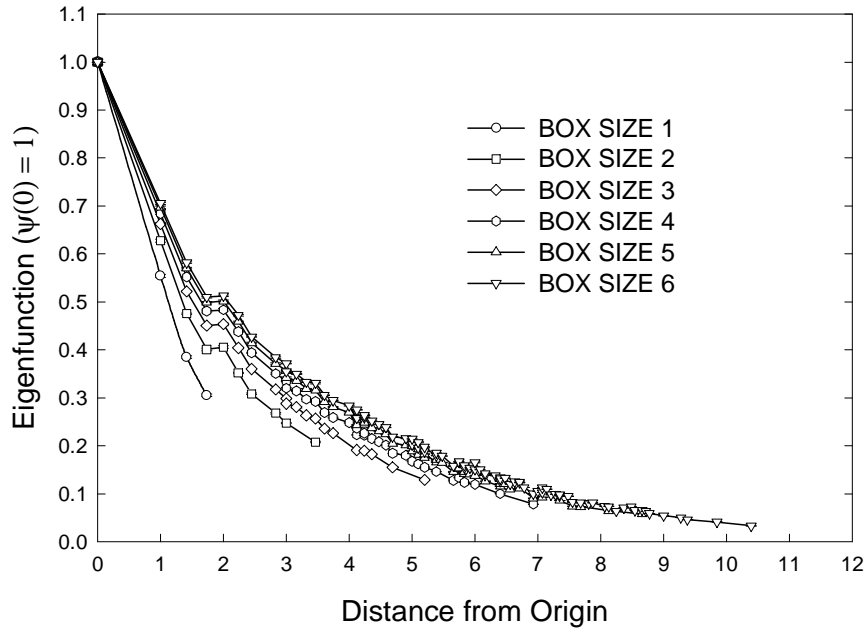

Figure 2 — The static optimal weighting function ("the heavy-light meson wave function") on domains with $(2N_W + 1) = 3, 5, 7, 9, 11$, and 13 sites on a side

The static component of the optimal weighting function on each of these domains is shown in Figure 2. Beyond the smallest domain, a box with only three sites on a side, its convergence and stability are evident. The glitch in each of the curves at distance 2 is a lattice artifact. It arises because the sites 2 units from the origin (*e.g.* (2,0,0)) are closer in lattice hops than those at a distance $\sqrt{3}$ units (*e.g.* (1,1,1)). The graphs are for the optimal weighting functions on time slice $t = 6$, but in fact the functions are very stable and their variation from one time slice to the next is less than 1%.

The higher terms in the expansion of the eigenvalue equation Eq. (24) are semi-negative-definite inhomogeneous equations. Suppressing the matrix labels ($\vec{y}, \vec{y}'$), the general equation is

$$\left[K_{[\vec{p}]}(t,\vec{m}=0) - \lambda_{[\vec{p}]}(t,\vec{m}=0)\right] \psi_{[\vec{p}]}(t,m_i) \\ = - \sum_{\vec{m}'} \left[K_{[\vec{p}]}(t,m_i') - \lambda_{[\vec{p}]}(t,m_i')\right] \psi_{[\vec{p}]}(t,m_i - m_i') \tag{31}$$

The sum goes over all $\vec{m}'$ satisfying $0 \leq m_i' \leq m_i$, but excluding $\vec{m}' = (0,0,0)$. These are actually a hierarchy of inhomogeneous equations for the expansion coefficients of both the eigenvalue and the eigenfunction. The equation for the $(m_1, m_2, m_3)$ coefficients is solved in terms of all the $(m_1', m_2', m_3')$ coefficients with $m_i' \leq m_i$, but subject to the strict equality $\sum m_i' < \sum m_i$. Note that even though $\psi_{[0]}(t,\vec{y},\vec{m}=0)$ is invariant under cubic transformations of $\vec{y}$, the higher, $\vec{m}' \neq 0$ terms are always asymmetric.

All of the inhomogeneous equations for varying $\vec{m}'$ have the same kernel,

$$\left[K_{[\vec{p}]}(t,\vec{y},\vec{y}',\vec{m}=0) - \lambda_{[\vec{p}]}(t,\vec{m}=0)\delta_{\vec{y},\vec{y}'}\right] = \\ e^{i\vec{p}\cdot\vec{y}}\left[K_{[0]}(t,\vec{y},\vec{y}',\vec{m}=0) - \lambda_{[0]}(t,\vec{m}=0)\delta_{\vec{y},\vec{y}'}\right]e^{-i\vec{p}\cdot\vec{y}'} \tag{32}$$

which, of course gives zero when applied to $\psi_{[\vec{p}]}(t,\vec{y},\vec{m}=0)$. The component of the equation along $\psi_{[\vec{p}]}(t,\vec{y},\vec{m}=0)$ gives the eigenvalue coefficient:

$$\lambda_{[\vec{p}]}(t,\vec{m}) = \sum_{\vec{m}' \neq 0}^{\vec{m}} \psi_{[\vec{p}]}^{\dagger}(t,\vec{m}=0) K_{[\vec{p}]}(t,\vec{m}') \psi_{[\vec{p}]}(t,\vec{m}-\vec{m}') \qquad (33)$$

The kernel restricted to the orthogonal subspace is negative definite and so can be inverted non-singularly on that space:

$$\psi_{[\vec{p}]}(t,\vec{m}) = -\left\{ Q_{[\vec{p}]} \left[ K_{[\vec{p}]}(t,\vec{m}=0) - \lambda_{[\vec{p}]}(t,\vec{m}=0) \right] Q_{[\vec{p}]} \right\}^{-1} \\ \times \sum_{\vec{m}' \neq 0}^{\vec{m}} \left[ K_{[\vec{p}]}(t,m_i') - \lambda_{[\vec{p}]}(t,m_i') \right] \psi_{[\vec{p}]}(t,m_i - m_i') \qquad (34)$$

where

$$Q_{[\vec{p}]}(\vec{y},\vec{y}') = e^{i\vec{p}\cdot\vec{y}} \left[ \delta_{\vec{y},\vec{y}'} - \psi_{[0]}(t,\vec{y},\vec{m}=0) \psi_{[0]}^{\dagger}(t,\vec{y}',\vec{m}=0) \right] e^{-i\vec{p}\cdot\vec{y}'} \qquad (35)$$

is the projector on the space orthogonal to the ground state. The matrix inverse is non-singular on the image space of $Q_{[\vec{p}]}$. Note that only a single matrix needs to be inverted throughout this process.

The successive terms in the expansion of the weighting function have different large Euclidean time behaviors. Specifically, the relative growth with $t$ of the terms in the expansion of the heavy quark propagator with different values of $\vec{m}$ (Eq. (9)) results in the same relative $t$ dependence of the terms in the expansion of the kernel (Eq. (23)). Since $t$ is a fixed parameter in the whole hierarchy of expressions for the wave function components (Eq. (34)), the coefficient $\psi_{[\vec{p}]}(t,\vec{y},\vec{m})$ grows with $t$ like

$$\psi_{[\vec{p}]}(t,\vec{y},\vec{m}) \sim t^{(m_1 + m_2 + m_3)} \tag{36}$$

This is the origin of the remark made earlier that while $\psi_{[\vec{p}]}(t,\vec{y},\vec{m}=0)$, which is asymptotically constant, can be interpreted as the wave function of a static composite meson, the full weighting function does not have that interpretation.

## V  SIMULATION OF $v^{(phys)}$

We have implemented these ideas on an ensemble of lattices and Wilson light quark propagators made available to us by the Fermilab ACP-MAPS Collaboration[17]. The ensemble consisted of 42 lattices of size $24^3 \times 48$ with lattice coupling $\beta = 6.1$ along with Wilson quark propagators with hopping constant $\kappa = .154$. Heavy quarks propagate in only one Euclidean time direction, but which direction is conventional. Each lattice therefore provides effectively two independent configurations, one each for association of the direction of time propagation of the heavy quark with the nominal forward or backward direction of lattice time.

The starting point for the optimization of the composite meson operator described in the previous section is the propagation kernel defined by Eq. (23), or more precisely its expansion coefficients $K_{[\vec{p}]}(t, \vec{y}, \vec{y}', \vec{m})$. These are the quantities that are the immediate output of a Monte Carlo simulation. To go from this starting point to the expansion coefficients of the physical classical velocity involves a long chain of operations that includes many non-linear steps. Because of this, a straightforward propagation of statistical errors is unlikely to give a valid estimate of the precision of the final result. We therefore follow a full single-elimination jackknife procedure, from the propagation kernel to the physical classical velocity, to compute the statistical precision of all results.

For each of the subensembles of all but one of the configurations, we evaluate the $\tilde{v}$ expansion coefficients in the propagation kernel defined by Eq. (23). We restrict the domain of the kernel to a box of relative coordinates with 5 sites on a side ($N_W = 2$). This is a compromise value. The choice $N_W = 3$ would yield a more rapid approach to a plateau, but would require a great deal more computer memory. The $N_W = 2$ calculation is sufficiently less demanding of memory to allow the simulation of all the terms in the $\tilde{v}$ expansions through third order. As we shall see, however, the third order coefficients are poorly determined, although the linear term is quite precisely computed.

The order of battle is as follows. We solve the zeroth order eigenvalue equation, Eq. (30) for the highest eigenvalue and its associated eigenvector with $\vec{p} = 0$. The eigenfunction on the 6$^{th}$ time slice is the second graph in Figure 2. There are 10 inequivalent sites, the furthest being $2\sqrt{3}$ from the origin. We then use Eq. (29) to obtain the zeroth order eigenvector for $\vec{p} = (1,0,0)$ and its permutations, and then iterate Eqs. (33) and (34) to find all the terms through third order, $m_1 + m_2 + m_3 \leq 3$. From the classical velocity expansion of Eq. (25), or equivalently of Eq. (22), we find the leading terms in the classical velocity expansion of the variationally optimized heavy-light meson propagator.

The zeroth and first order terms in the expansion of the heavy-light meson propagator $M^{(v)}$ that enter into the determination of the physical classical velocity are shown in Figure 3. The values of these and all the second and third order components of the propagator that enter Eq. (20) are tabulated in Table I. Note that since the components with $m_i$ odd are odd in $p_i$ as well, the values given for $M(t,\vec{p}=(1..),\vec{m}=(odd..))$ are exactly the same as $p_{min} = 2\pi/Na$ times the lattice approximation to the momentum derivative $\Delta M/\Delta p_i|_{\vec{p}=0}$ of Eq. (18). The precision of the leading terms and their convergence to the expected asymptotic forms is striking. The error bars on the computed propagator values are from the single-elimination jackknife analysis, and the error bars on the percentage saturation are the fractional errors on the static propagator. The fit to the static propagator is a least squares two-exponential fit to the computed values:

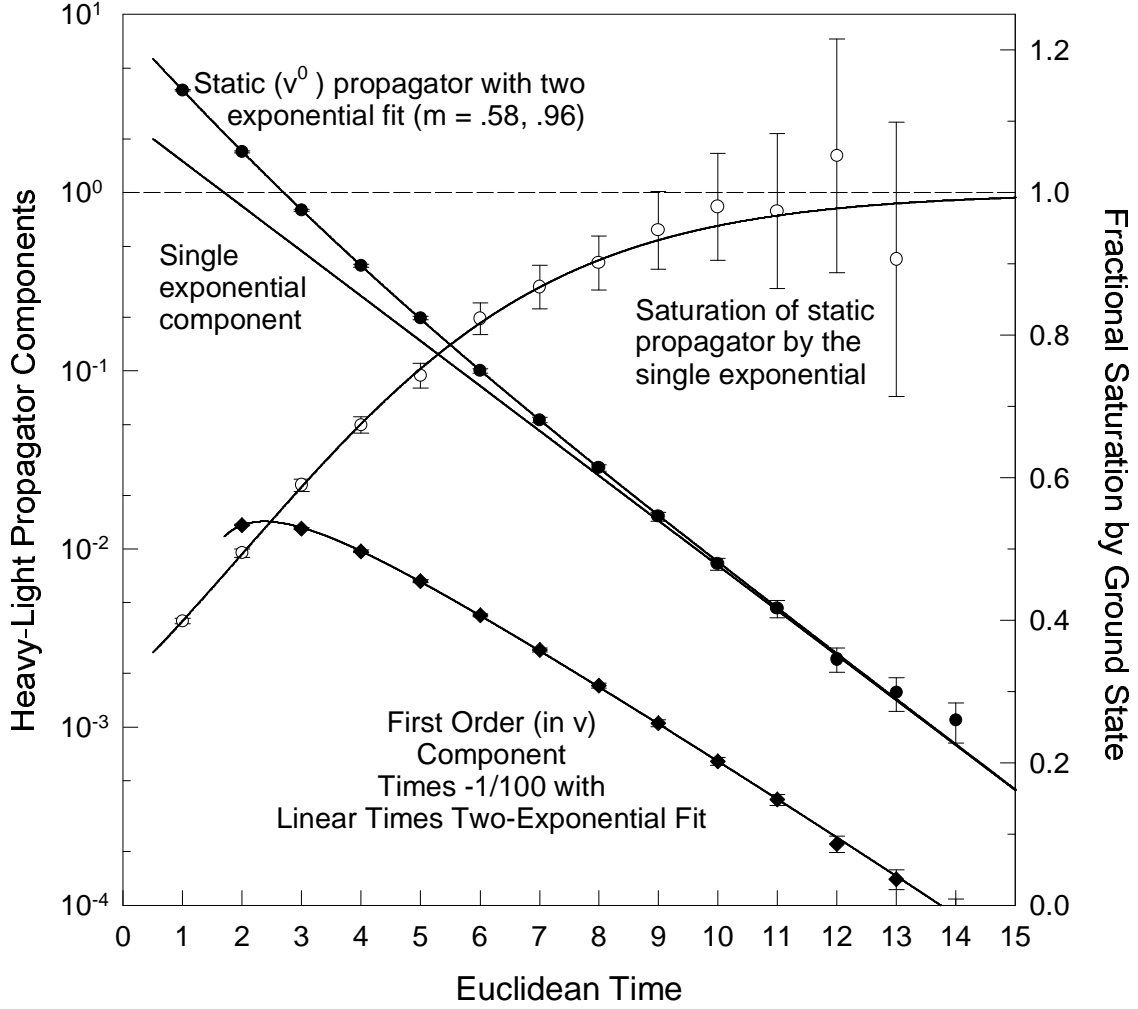

Figure 3 — The zeroth and first order components of the optimized heavy-light meson propagator, showing degree of dominance by the ground state

$$M(t, \vec{p}=0, \vec{m}=0) \sim c_1 e^{-m_1 t} + c_2 e^{-m_2 t} \tag{37}$$

The masses are $m_1 = .580 \pm .012$ and $m_2 = .961 \pm .031$, and the ratio of the coefficients is $c_2/c_1 = 2.20 \pm .13$. The higher mass term effectively models all of the higher mass effects in the static propagator. The simple exponential plotted under the static propagator is the first term from the above fit. The asymptotic mass is, of course, very slightly lower than that obtained from the time-slice to time-slice fall-off the propagator below $t = 10$. The computed saturation values are the single exponential component of the two-exponential fit divided by the propagator values, and the saturation curve is the ratio of the single exponential to that fit. The curve through the computed values of the first order component of the meson propagator is a least squares fit to a linear function of lattice time times the two-exponential fit to the static propagator, using the values of the coefficients $c_i$ and $m_i$ determined from the static propagator

$$M(t,\vec{p}=(100),\vec{m}=(100)) \sim (a + bt)\left(c_1 e^{-m_1 t} + c_2 e^{-m_2 t}\right) \qquad (38)$$

Table I — Coefficients $M(t,\vec{p},\vec{m})$ in the Optimized Composite Meson Propagator Expansion
All terms are averaged over equivalent directions

| Time Slice | $\vec{m} = (0,0,0)$ $\vec{p} = (0,0,0)$ | $\vec{m} = (1,0,0)$ $\vec{p} = (1,0,0)$ | $\vec{m} = (2,0,0)$ $\vec{p} = (0,0,0)$ | $\vec{m} = (3,0,0)$ $\vec{p} = (1,0,0)$ | $\vec{m} = (1,2,0)$ $\vec{p} = (1,0,0)$ |
|---|---|---|---|---|---|
| 1  | 3.7459 ± .0356 |                |                 |                 |                 |
| 2  | 1.6924 ± .0204 | -1.3645 ± .0168 |                |                 |                 |
| 3  | 0.7940 ± .0116 | -1.3011 ± .0186 | -0.1175 ± .0054 |                 |                 |
| 4  | 0.3888 ± .0065 | -0.9687 ± .0154 | -0.0410 ± .0052 | 0.3727 ± .0132  | 0.4353 ± .0152  |
| 5  | 0.1974 ± .0046 | -0.6589 ± .0132 | 0.0044 ± .0047  | 0.2766 ± .0174  | 0.3551 ± .0216  |
| 6  | 0.0998 ± .0027 | -0.4240 ± .0092 | 0.0145 ± .0047  | 0.1406 ± .0212  | 0.2123 ± .0261  |
| 7  | 0.0530 ± .0019 | -0.2714 ± .0076 | 0.0162 ± .0036  | 0.0367 ± .0207  | 0.0946 ± .0255  |
| 8  | 0.0286 ± .0012 | -0.1713 ± .0058 | 0.0128 ± .0034  | -0.0110 ± .0204 | 0.0363 ± .0230  |
| 9  | 0.0152 ± .0009 | -0.1053 ± .0046 | 0.0075 ± .0040  | -0.0250 ± .0218 | 0.0091 ± .0253  |
| 10 | 0.0082 ± .0006 | -0.0643 ± .0034 | 0.0049 ± .0041  | -0.0249 ± .0221 | 0.0027 ± .0260  |
| 11 | 0.0046 ± .0005 | -0.0391 ± .0027 | 0.0068 ± .0039  | -0.0056 ± .0335 | -0.0087 ± .0345 |
| 12 | 0.0024 ± .0004 | -0.0221 ± .0023 | -0.0019 ± .0054 | 0.0525 ± .0622  | 0.0129 ± .0566  |
| 13 | 0.0016 ± .0003 | -0.0140 ± .0018 | -0.0070 ± .0071 | 0.0403 ± .0648  | 0.0328 ± .0666  |
| 14 | 0.0011 ± .0003 | -0.0091 ± .0018 | -0.0046 ± .0065 | 0.1009 ± .0825  | 0.1040 ± .0845  |
| 15 | 0.0005 ± .0002 | -0.0047 ± .0016 | -0.0085 ± .0079 | 0.0797 ± .0806  | 0.0603 ± .0796  |

This form is chosen to agree with the expected asymptotic behavior.

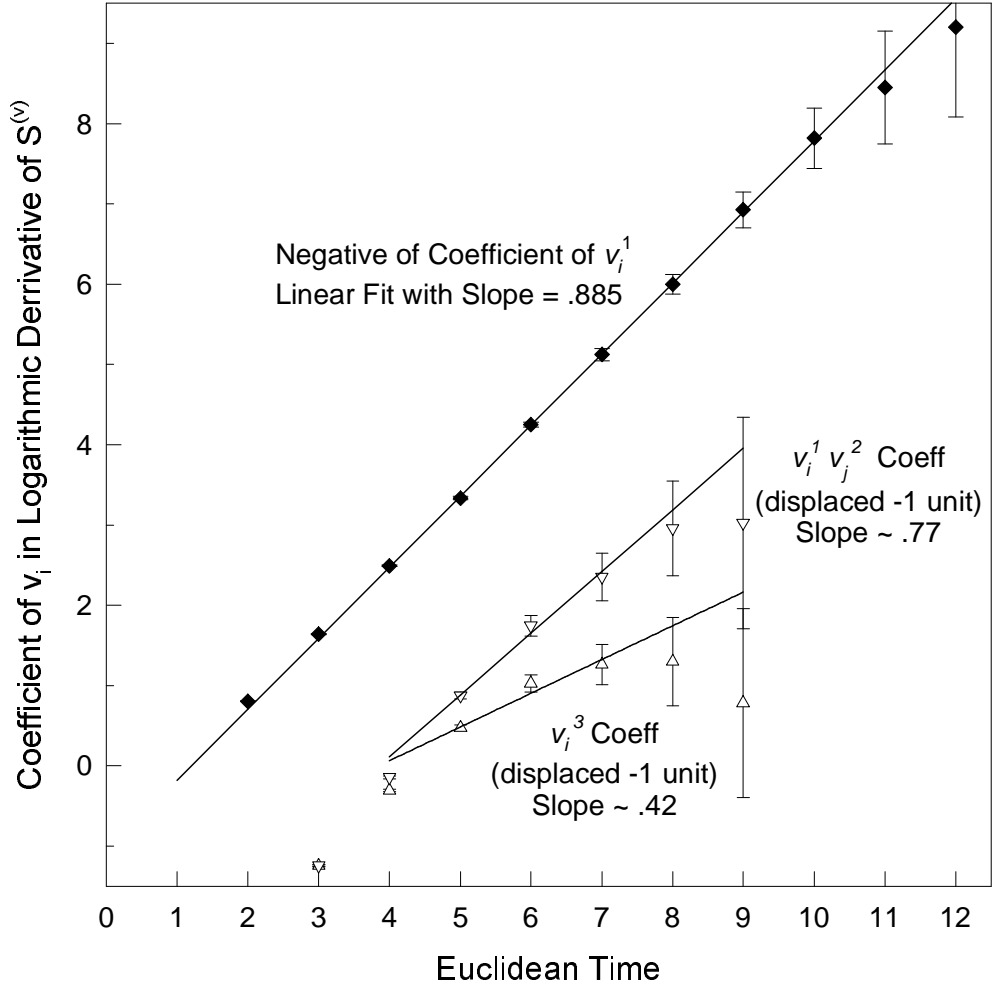

Figure 4 — $\tilde{v}_i$ Expansion Functions of the Logarithmic Derivative of the Meson Propagator

The functions of Eq, (20), whose slopes determine the physical classical velocity up to third order in the input classical velocity are given for the optimized composite meson propagator by combinations of the components of Table I. All three are plotted in Figure 4 and are also given in Table II. The indicated statistical errors are not propagated from the errors in the individual components but directly obtained from a single-elimination jackknife procedure.

Table II — Leading Terms in the Expansion of the Logarithmic Derivative of the Optimized Meson Propagator ( averaged over equivalent directions)

| Time Slice | $R^{(i)}(m_i=1)$ | $R^{(i)}(m_i=3)$ | $R^{(i)}(m_i=1, m_{j\neq i}=2)$ |
|---|---|---|---|
| 1 | | | |
| 2 | -0.806 ± .005 | | |
| 3 | -1.639 ± .007 | -0.242 ± .011 | -0.242 ± .011 |
| 4 | -2.491 ± .010 | 0.696 ± .014 | 0.857 ± .017 |
| 5 | -3.338 ± .021 | 1.476 ± .035 | 1.874 ± .043 |
| 6 | -4.249 ± .031 | 2.026 ± .108 | 2.745 ± .129 |
| 7 | -5.122 ± .074 | 2.263 ± .251 | 3.355 ± .297 |
| 8 | -5.998 ± .123 | 2.298 ± .548 | 3.957 ± .588 |
| 9 | -6.926 ± .224 | 1.783 ±1.174 | 4.027 ±1.318 |
| 10 | -7.819 ± .376 | | |
| 11 | -8.451 ± .702 | | |
| 12 | -9.203 ±1.121 | | |

The "asymptotic" linearity of the leading, first order term in the logarithmic derivative of the meson propagator is evident from the first time slice on which it is non-zero, $t = 2$. The slope of the linear fit plotted, .885, is the average of the best fits over the intervals $t = 4$ - 10 ($-.873 \pm .011$) and $t = 5$ - 9 ($-.897 \pm .023$). The difference between the two fitted slope values is driven by the value of $R^{(i)}(m_i=1)$ on the $t = 4$ time slice. It has the smallest fractional error, but might not be quantitatively asymptotic. The average of their statistical errors, which is ± .017, is larger than half the difference between the values of the slope determined on the above two intervals, ± .012, and is a conservative estimate of the statistical precision of the fitted slope.

The quality of the determinations of the slopes of the two combinations of propagators that give the third-order coefficients are not at all as satisfactory. The linear fits plotted in

Table III — The linear and cubic terms in the expansion of the physical classical velocity in powers of the input classical velocity

| Coefficient (Cf. Eq. (17)) | Multi-plies | Fitted Value (Statistical Error) | Lattice Time Interval for Fit |
|---|---|---|---|
| $c^{(i)}(m_i = 1)$ | $\tilde{v}_i$ | $.885 \pm .017$ | Average of 4 - 10 and 5 - 9 |
| $c^{(i)}(m_i = 3)$ | $\tilde{v}_i^3$ | $-.420 \pm .082$ | 5 - 9 |
| $c^{(i)}(m_i = 1, m_{j \neq i} = 2)$ | $\tilde{v}_i \tilde{v}_j^2$ $(i \neq j)$ | $-.770 \pm .058$ | 5 - 9 |

Figure 4 are least squares straight line fits on the interval $t = 5$ - $9$. The slopes and their jackknife statistical errors are given Table III. It is not the statistical precision of these fits which is at issue, but the fact that the statistical precision of the computed values of the propagator ratios being fit deteriorates so quickly with lattice time that one must admit the possibility that the true asymptotic linear behavior has not been seen at all in this simulation.

The fact that the two third order coefficients are not equal is due to the violation of rotational invariance by the lattice. Continuum rotations would transform the tensor components $\tilde{v}_i^3$ and $\tilde{v}_i \tilde{v}_j^2$ into linear combinations of each other, but lattice rotations do not mix them. As we remarked in the Introduction however, these coefficients would vanish entirely in the continuum, so their inequality is not by itself a reason to distrust the fitted slopes.

While it is true that each of the (asymptotically linear) third order functions is the difference of terms growing more rapidly with lattice time ($\propto t^3$), it is nonetheless disappointing that the variationally optimized meson propagator is not sufficiently robust to unambiguously extract these functions.

## VI  ONE-LOOP PERTURBATIVE RENORMALIZATION OF $v$

The perturbative calculation of the renormalization of the classical velocity is phrased in terms of the proper self mass $\Sigma^{(v)}(p)$, which is related to the propagator by

$$S^{(v)}(p) = \frac{1}{v \cdot p - \Sigma^{(v)}(p)} \tag{39}$$

For the purpose of identifying the physical classical velocity, it is sufficient to examine only the leading terms in the Taylor expansion of $\Sigma^{(v)}(p)$ in powers of the residual momentum:

$$\Sigma^{(v)}(p) = m + X_\mu p_\mu + \cdots$$

$$X_\mu = \left.\frac{\partial \Sigma^{(v)}}{\partial p_\mu}\right|_{p=0} \tag{40}$$

The first term is a mass shift. This is without physical significance because it can be eliminated by a redefinition of the reduced heavy quark field by a non-singular phase which is independent of the heavy mass[18].

$$h^{(v)}(x) \to e^{imv \cdot x} h^{(v)}(x) \tag{41}$$

This corresponds to a finitely different breakup of the total 4-momentum

$$P = Mv + p \to Mv' + (p + mv) \tag{42}$$

In the heavy quark limit, $v$ and $v'$ become equal. We will drop any possible residual mass term from the remaining discussion.

The renormalized classical velocity and the wave function renormalization are both inferred from the small residual momentum behavior of the full heavy quark propagator

$$S^{(v)}(p) = \frac{Z^{(v)}}{v^{(ren)} \cdot p + \cdots} \tag{43}$$

The $p \to 0$ behavior of the propagator is sufficient to obtain both renormalizations because the bare and the normalized classical velocities are separately normalized to $-1$.

$$v_\mu^{(ren)\,2} = v_\mu^2 = Z^{(v)\,2}(v_\mu - X_\mu)^2 = -1 \tag{44}$$

To lowest order in perturbation theory, the renormalized bounded classical velocity is given by

$$\tilde{v}_i^{(ren)} \equiv \frac{v_i^{(ren)}}{v_0^{(ren)}} = \tilde{v}_i - \frac{1}{v_0}(X_i - X_0 \tilde{v}_i) \tag{45}$$

As was discussed in the Introduction, if $X_\mu$ is proportional to $v_\mu$, which is always the case in the continuum, the classical velocity is not renormalized.

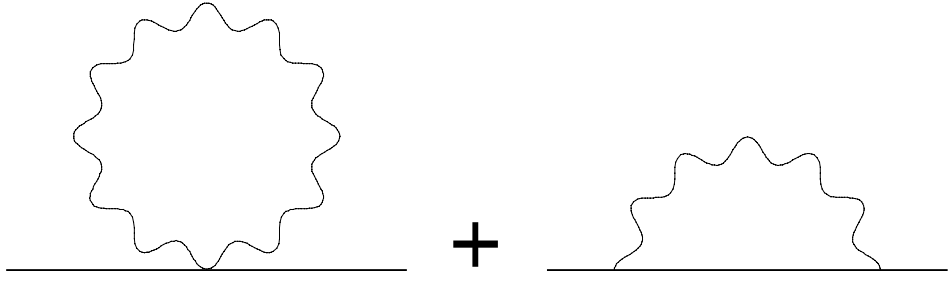

Figure 5 — The 1-loop contribution to the heavy quark proper self mass

The two diagrams that contribute to the proper self mass at one loop are shown in Figure 5. The point interaction only gives rise to a residual mass, which we will ignore. With an infrared regulator $\lambda$, the second diagram gives

$$\Sigma^{(2)} = g^2 C_2 \int \frac{d^4k}{(2\pi)^4} V_\mu(p-2k) \Delta_{(\lambda)}(k) S^{(v)}(p-k) V_\mu(p-2k) \qquad (46)$$

The integration domain is periodic, $[-\pi/a, \pi/a]$, and the bare propagators and vertices are

$$S^{(v)-1}(p) = \frac{v_0}{a}\left(e^{ip_4 a} - 1\right) + \sum_i \frac{v_i}{a} \sin p_i a$$

$$\Delta_{(\lambda)}^{-1}(k) = \frac{4}{a^2} \sum_\mu \sin^2 \frac{k_\mu a}{2} + \lambda^2 \qquad (47)$$

$$V_\mu^a(q = 2p+k) = g t^a \left(i v_0 e^{iqa/2},\ v_i \cos \frac{q_i a}{2}\right)$$

The factor of the quadratic Casimir invariant, $C_2$, which is $(N^2-1)/2N$ for SU($N$) color with quarks in the fundamental representation, arises from the color contraction in the loop integral.

The Taylor series coefficient of $p_\mu^1$ is a non-linear function of $v_\mu$ and is furthermore asymmetric between space and time (because of the asymmetry of the lattice action, which is a centered difference in space but a forward difference in time). Thus perturbatively there is also a classical velocity shift as well as a multiplicative renormalization constant $Z^{(v)}$.

The evaluation of the loop integral over the gluon 4-momentum requires care in the choice of contour in the complex Euclidean energy plane. The path of the contour with respect to the poles in the propagators is determined by the coordinate space boundary conditions. The contour must be chosen so that the heavy quark propagator always vanishes for negative Euclidean time. Thus the propagator as a function of momentum and Euclidean time is given by

$$\int \frac{dE}{2\pi} \frac{e^{iEt}}{\frac{v_0}{a}[e^{iEa} - 1] + \frac{1}{a}\sum_i v_i \sin p_i a}$$

$$= \frac{1}{2\pi i} \oint \frac{dz}{z} \frac{z^{(t/a)}}{v_0[z - 1] + \sum_i v_i \sin p_i a} \tag{48}$$

$$(z = e^{iEa})$$

and the necessity for this to vanish for negative $t$ no matter the values of $v$ and $p$ implies that the contour in the $z$ plane always encloses the pole.

Although 3-momentum integrals will eventually have to be done numerically, it is necessary to do the energy integration analytically. With the notation

$$z = e^{iak_4}$$

$$A = 4 \sum_{i=1}^{3} \sin^2 \frac{k_i a}{2} + (\lambda a)^2$$

$$B = \sum_{i=1}^{3} v_i \sin(p_i + k_i) a \tag{49}$$

$$C = \sum_{i=1}^{3} v_i^2 \cos^2 \left( p_i + \frac{k_i}{2} \right) a$$

the Euclidean energy integral is

$$\frac{a^2}{2\pi i} \oint \frac{dz}{z} \frac{- v_0^2 e^{2ip_4 a} z + C}{[-(z - 2 + 1/z) + A][v_0 (e^{ip_4 a} z - 1) + B]} \tag{50}$$

where the integration is around a closed contour in the $z$ plane. Note that while the quantities $A$ and $C$ are both positive, $B$ can take either sign. The singularities of the integrand are located at

$$z_{\pm} = 1 + \frac{A}{2} \pm \frac{1}{2} \sqrt{4A + A^2}$$

$$z_q = e^{-ip_4 a} \left( 1 - \frac{1}{v_0} B \right) \tag{51}$$

One of the pair of poles coming from the gluon propagator is within the unit circle in the $z$ plane, and the other is outside. Depending on the sign of $B$, the pole coming from the heavy quark propagator can be either inside or outside. The contour must be chosen to pass between the two gluon propagator poles, which is the ordinary procedure. However, the requirement that the heavy quark propagator must vanish for all negative Euclidean times means that the contour must always enclose the quark propagator pole, whatever the sign of $B$[19]. The appropriate contour is shown in Figure 6, for the non-standard case, negative $B$.

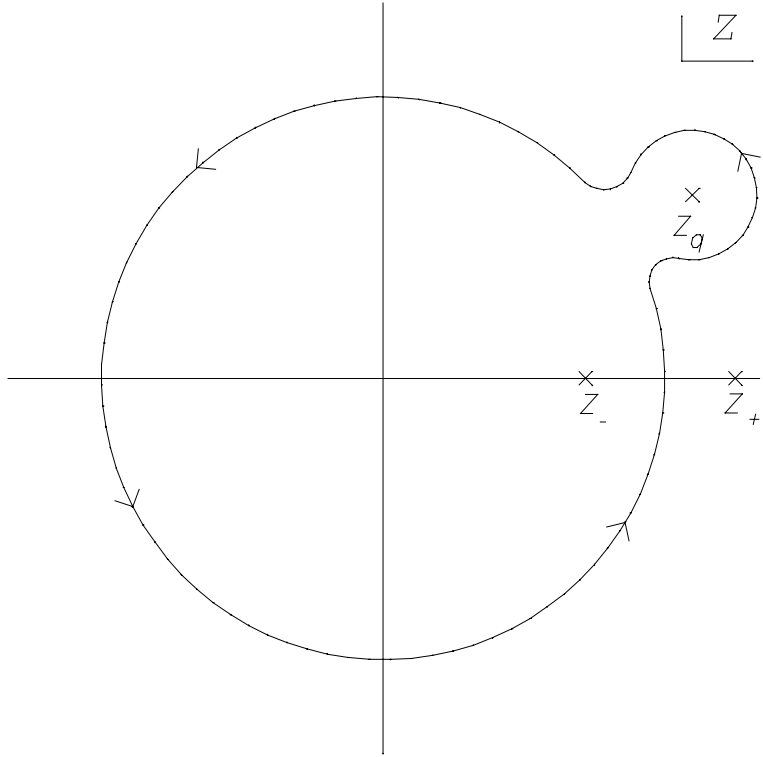

Figure 6 — Contour in the $z = e^{iEa}$ plane for the one loop proper self mass

The integrand decays sufficiently rapidly at infinity so that the energy integral is given by the residue of the single pole outside the contour, at $z_+$. The resulting expression for the one loop proper self mass is

$$\Sigma^{(2)} = g^2 C_2 \frac{a^2}{(2\pi)^3} \int d^3k \frac{-v_0^2 e^{2ip_4 a} z_+ + C}{\sqrt{4A + A^2} \, [v_0 (e^{ip_4 a} z_+ - 1) + B]} \tag{52}$$

To facilitate comparison with the simulation of the physical classical velocity, we take the derivative with respect to $p_\mu$ at the origin, which determines the renormalized classical velocity, and then expand the bounded classical velocity shift in powers of the input classical velocity

$$\delta \tilde{v}_i = [c^{(i)}(m_i=1) - 1] \tilde{v}_i + c^{(i)}(m_i=3) \tilde{v}_i^3 + \sum_{j \neq i} c^{(i)}(m_i=1, m_j=2) \tilde{v}_i \tilde{v}_j^2 + \cdots \tag{53}$$

The linear and cubic expansion coefficients are given by

$$c^{(i)}(m_i=1) - 1 = \frac{-g^2 C_2}{(2\pi)^3} \int d^3\theta \frac{z_+ (\cos\theta_i + z_+ - 2)}{\sqrt{4A + A^2} \, (z_+ - 1)^2}$$

$$c^{(i)}(m_i=3) = \frac{-g^2 C_2}{(2\pi)^3} \int d^3\theta \frac{(\cos\theta_i + 1)(z_\pm 3\cos\theta_i + 2)(z_+^2 + 2z_+(\cos\theta_i - 2) + 1)}{2\sqrt{4A + A^2} \, (z_+ - 1)^4} \tag{54}$$

$$c^{(i)}(m_i=1, m_j=2) = \frac{-g^2 C_2}{(2\pi)^3} \int d^3\theta \frac{(\cos\theta_j + 1) \begin{bmatrix} 2z_+(z_+ - 3\cos\theta_i + 2)\cos\theta_j - \\ (z_+^2 - 8z_+ + 1)\cos\theta_i + z_+(z_+^2 - 4z_+ - 3) \end{bmatrix}}{2\sqrt{4A + A^2} \, (z_+ - 1)^4}$$
$$(j \neq i)$$

where the integration is over the periodic box $\theta_i \in [-\pi, \pi]$.

The numerical values of these coefficients, for $g^2 = 6/\beta = 6/6.1$ are

$$c^{(i)}(m_i=1) - 1 = -.23345569$$
$$c^{(i)}(m_i=3) = -.04143250 \tag{55}$$
$$c^{(i)}(m_i=1, m_j=2) = -.03881061 \quad (i \neq j)$$

The difference between the values of the two cubic terms is a reflection of the reduced spatial symmetry of the lattice.

Table IV — Comparison of Simulation with Perturbation Theory Classical Velocity Shifts

| Coefficient | Value from Simulation | Perturbative Evaluation |
|---|---|---|
| $c^{(i)}(m_i = 1) - 1$ | $-.115 \pm .017$ | $-.23345569$ |
| $c^{(i)}(m_i = 3)$ | $-.420 \pm .082$ | $-.04143250$ |
| $c^{(i)}(m_i = 1, m_{j \neq i} = 2)$ | $-.770 \pm .058$ | $-.03881061$ |

It is striking to compare the perturbative evaluation of these expansion coefficients with the results of the simulation performed at $\beta = 6.1$. This is summarized in Table IV. There is a rather complete disagreement between the two sets of coefficients. The linear terms disagree by a factor of 2. The simulation gives cubic terms order 1, while in perturbation theory they are tiny. The clear linearity (versus lattice time, Figure 4) of the ratio of the simulated propagator components from which the leading coefficient is determined argues that the simulated value is thoroughly stable. That is not true for the third order coefficients, however. We regard it as quite plausible that a better simulation could result in totally different values for these coefficients. The complete disagreement between the simulation and perturbative first order terms certainly undermines confidence that perturbation theory is a trustworthy method of calculating the classical velocity shift.

Aglietti and Giménez have carried out an extensive one-loop calculation of the renormalization of the lattice HQET, including but not limited to the classical velocity renormalization[13]. However, in place of the discretization of the Dirac equation used here, Eq. (6), their discretization used an asymmetric backward time difference rather than the asymmetric forward time difference we have used in this analysis:

$$[U_4(x,x+\hat{t}) S^{(v)}(x+\hat{t},y) - S^{(v)}(x,y)] \Rightarrow [S^{(v)}(x,y) - U_4(x,x-\hat{t}) S^{(v)}(x-\hat{t},y)]^{(AG)} \quad (56)$$

While the two are obviously equal in the continuum, and both are suitable for perturbative calculations, the Aglietti-Gimémez form of the Dirac equation is much less convenient for simulations. The reason is that instead of a few matrix operations, it requires the solution of a difference equation over a full time slice for each step in time. This modification of the Dirac equation results in two alterations of the lattice Feynman rules given in Eq. (47):

$$\begin{aligned} S^{(v)^{-1}}(p): & \left[e^{ip_4 a} - 1\right] \Rightarrow \left[1 - e^{-ip_4 a}\right]^{(AG)} \\ V_4^a(q): & \left[e^{iqa/2}\right] \Rightarrow \left[e^{-iqa/2}\right]^{(AG)} \end{aligned} \quad (57)$$

This modifies the precise location in the complex energy plane of the pole coming from the quark propagator, but the general structure of the contour integral is unchanged.

The renormalization of the classical velocity is qualitatively different from the renormalizations of quantities that match to continuum values. It has no divergent part, and in fact it vanishes in the continuum. Accordingly, one might suspect that it will depend crucially on the precise form of the lattice discretization. This is in fact the case. For $g^2 = 6/\beta = 6/6.1$, the numerical values of the coefficients computed above, using the Aglietti-Giménez rules, are

$$c^{(i)}(m_i=1)^{[AG]} - 1 = + .09799480$$
$$c^{(i)}(m_i=3)^{[AG]} = + .00282556 \tag{58}$$
$$c^{(i)}(m_i=1, m_j=2)^{[AG]} = + .05811408 \quad (i \neq j)$$

The first two coefficients were previously calculated in Ref. [13]. The third arises only when more than one component of the classical velocity in non-zero, a case not treated in [13]. The two sets of coefficients obtained from the different discretizations of the heavy quark Dirac equation are completely different; they even have opposite signs. This is a clear sign that the rescaling of the classical velocity is quite different in character from other, more familiar lattice renormalizations, which are analogues of the continuum renormalizations that remove divergences.

A second way in which the rescaling of the classical velocity is quite different from other lattice/continuum matching processes is that it is not improved by "Tadpole Improvement"[20]. In fact, the application of Lepage and Mackenzie's tadpole improvement procedure worsens the disagreement between the perturbative calculation of the classical velocity renormalization and the results of the simulation. Quantitatively, the principal effect of tadpole improvement in the present context comes from the reidentification of the couplings used in the simulation with that used in the perturbative analysis. To cancel the lattice tadpole corrections to the gauge field, one replaces each link variable $U_\mu(n)$ by that link divided by its average value $U_\mu(n)/u_0$, where the average value of the link is given by the gauge-invariant expression

$$u_0 = \left[\frac{1}{3}\langle Tr \Box \rangle\right]^{1/4} \tag{59}$$

and $\Box$ denotes the product of the link variables around an elementary 1×1 plaquette. For the Wilson action used to generate the Fermilab lattices, this effects the change

$$I = -\frac{6}{g^2}\sum_\Box \frac{1}{3} Tr \Box \Rightarrow -\frac{6}{g^2 u_0^4}\sum_\Box \frac{1}{3} Tr \Box \tag{60}$$

and so β is identified with $6/g^2 u_0^4$ rather than $6/g^2$. Since $u_0$ is necessarily less than one, this always has the effect of increasing the value of the effective perturbative coupling corresponding to a lattice simulation at a given value of β. The value of $u_0^4$ at β = 6.1 is about 0.6, which would have the effect of turning the factor of 2 discrepancy into a factor of 3 discrepancy.

## VII    CONCLUDING REMARKS

Let us sum up the results of these analyses of the renormalization of the classical velocity in lattice version of the heavy quark effective theory. The origin of this new renormalization is the reduction of the Lorentz invariance of the continuum to hypercubic symmetry on the lattice. Of the several perturbative calculation and Monte Carlo simulations we have presented, the most reliable result is most likely the simulation of the first-order, multiplicative shift in the classical velocity. The numerically computed propagator ratio $R^{(i)}(t, m_i=1)$ plotted in Figure 4 follows its theoretically expected behavior, Eqs. (15) and (17), quite accurately over an extended region of Euclidean time. For small classical velocities $\tilde{v}$ we have

$$\tilde{v}_i^{(phys)} \; = \; (.885 \pm .017)\, \tilde{v}_i \; + \; O(\tilde{v}_i^3, \tilde{v}_i \tilde{v}_j^2) \; + \; \cdots \tag{61}$$

We do not have the same level of confidence in the present computation of the third-order coefficients, Table III, that we have in the simulation of the linear shift.

We have no special insight into why the one-loop perturbative calculations of these coefficients differ so strongly from the results of the simulation, nor into why the Lepage-Mackenzie procedure worsens rather than improves the disagreement. It is of course easy to blame the fact that the lattice coupling corresponds to a continuum coupling constant near 1, but without at least a two-loop calculation of these coefficients, the matter will remain moot.

Finally, let us note that the renormalization of the classical velocity on the lattice affects the results of simulations of the Isgur-Wise function in the lattice HQET. Conveniently, in the light of the results of the present paper, it is only the linear, multiplicative shift that enters into the phenomenologically most important quantity, the slope of the Isgur-Wise form factor at the kinematic origin, $v \cdot v' = 1$. The cubic and higher order terms only affect its higher derivatives. Expressing the standard kinematic variable $v \cdot v' - 1$ in terms of the bounded classical velocity $\tilde{v}$

$$v \cdot v' - 1 \approx \frac{1}{2}(\tilde{v} - \tilde{v}')^2 + \frac{1}{4}(\tilde{v} - \tilde{v}')^2(\tilde{v}^2 + \tilde{v}'^2) + \frac{1}{8}(\tilde{v}^2 - \tilde{v}'^2)^2 + \cdots \quad (62)$$

we see that it is second order in $\tilde{v}$. Thus the rescaling $\tilde{v} \to \tilde{v}^{(phys)}$ has the effect of increasing the computed slope of the form factor with respect to $v \cdot v'$ at the origin by the square of the first order rescaling coefficient,

$$\frac{1}{c^{(i)}(m_i=1)^2} \approx 1.28 \pm .05 \quad (63)$$

**ACKNOWLEDGEMENT**

The authors would like to thank the Fermilab ACP-MAPS group for generously making available the lattice configurations and propagators that were used in the evaluation of the renormalization of the classical velocity.